# Ion energy measurements on MAST using a midplane RFEA

S. Y. Allan[1*], S. Elmore[1,2], A. Kirk[1], M. Kočan[3] and P. Tamain[4]

[1]*EURATOM/CCFE Fusion Association, Culham Science Centre, Abingdon, OX14 3DB, United Kingdom.*

[2]*The University of Liverpool, Brownlow Hill, Liverpool, L69 3GJ, United Kingdom.*

[3]*Max-Planck-Institut für Plasmphysick, Boltzmannstrasse, 2, D-85748, Garching, Germany.*

[4]*CEA, IRFM, F-13108, Saint-Paul-lez-Durance, France.*

**Abstract**

Ion energy measurements have been made in the scrape off layer of the Mega Amp Spherical Tokamak (MAST) using a midplane retarding field energy analyser (RFEA) in H-mode plasmas during the inter-edge localised mode (ELM) period and during type I and type III ELMs. During the inter-ELM period at distances of 3 to 8 cm from the last closed flux surface (LCFS), ion temperatures of 20 to 70 eV have been measured giving an ion to electron temperature ratio of 2 to 7 with a mean of 4. During type III ELMs, an ion temperature of 50 eV has been measured 3 to 6 cm from the LCFS which decreases to 30 eV at distances 11 to 16 cm from the LCFS. During type I ELMs, an ion temperature of 40 eV has been measured at a distance of 10 to 15 cm from the LCFS.

*PACS*: 52.70.Ds, 52.55.Fa

# 1. Introduction

One of the most important parameters in determining the extent of wall damage in tokamaks is the energy of ions impacting on the surface. Ion temperature determines sputtering yields and the quantity of impurities injected back into the plasma. Some of the most significant damage to the first wall occurs during ELMs where bursts of core plasma are released into the scrape off layer (SOL). The measurement of ion energies during both steady state and transient events such as ELMs is important for determining sputtering yields and for the development of particle and energy transport models of the scrape off layer where experimental measurements can provide boundary conditions for computer simulations.

One of the most widely used methods of measuring ion temperature is using an electrical probe called a Retarding Field Energy Analyser (RFEA) [1]. RFEAs have been used in a number of tokamaks including Alcator C [2], JET [3,4], ASDEX Upgrade [5], Tore Supra [6] and MAST [7,8]. Measurements in a number of tokamaks [9] have shown that the ratio of the ion temperature ($T_i$) to electron temperature ($T_e$) in the SOL of L-mode plasmas can range from 1 to 10. Ion energy measurements during ELMs have been made on ASDEX Upgrade [10] showing energies between 20 and 200 eV, 3.5 to 6 cm from the LCFS and during inter-ELM periods at approximately 5 cm from the LCFS, $T_i$ was found to be approximately 12 eV [5]. Measurements on JET during ELMs have measured ions with energies in excess of 400 eV, 5 cm from the LCFS [3]. Preliminary measurements on MAST during ELMs have measured ion energies between 200 and 500 eV at distances of 10 to 20 cm from the LCFS [7] but no measurement of ion temperature was possible.

In this paper, the results of inter-ELM and ELM $T_i$ measurements made on MAST [11] using a RFEA mounted on a reciprocating probe at the midplane will be presented. In section 2, a brief introduction to RFEAs is given along with a description of the plasma conditions used. In section 3, the results of inter-ELM and ELM measurements are presented.

## 2. Method

### 2.1 Retarding field energy analysers

A standard RFEA consists of a series of metal grids inside a probe body. Plasma enters the RFEA through a slit in the electrically grounded shell. Electrons are repelled by a negatively biased slit plate and ions are discriminated using a swept positive voltage on grid 1. The RFEA is aligned with its grid faces perpendicular to the magnetic field so that ions are discriminated based on the component of their velocity parallel to the magnetic field lines. The ion current is measured at a metal collector at the rear of the RFEA as a function of the grid 1 voltage. Grid 2 is biased negatively to minimise the effects of secondary electron emission on ion current measurements. For a given discriminator voltage, the collector current ($I_{col}$) is a measure of the ion velocity distribution function parallel to the magnetic field [1]. Assuming that the energy distribution of the ions along the magnetic field lines is described by a Maxwellian distribution, the effective ion temperature ($T_i^*$) can be determined by fitting a graph of $I_{col}$ versus grid 1 voltage ($V_{gr1}$) to equation (1):

$$I_{col} = I_o \exp\left[-\frac{Z_i}{T_i^*}\left(V_{gr1} - |V_s|\right)\right] + I_{off} \quad .....(1)$$

where $I_0$ is the ion saturation current, $Z_i$ is the effective ion charge (this work assumes $Z_i = 1$), $V_s$ is the plasma sheath voltage and $I_{off}$ is an offset current [12].

## 2.2 MAST RFEAs

The measurements presented here were performed in a range of H-mode deuterium plasmas using a RFEA located at the midplane on a reciprocating probe (RP). MAST is also fitted with a RFEA at the divertor target and $T_i$ measurements in H-mode from this RFEA are reported elsewhere [13]. The RP RFEA has two identical grid stack modules mounted in an electrically grounded graphite shell allowing bidirectional measurements of $T_i$. More detailed information on the design of the RFEAs and their electronics and the results of $T_i$ measurements in L-mode plasmas using both the RP and divertor RFEAs can be found in reference [8].

## 3. Results

## 3.1 Inter-ELM Measurements

Inter-ELM measurements were obtained in a double null H-mode discharge with 3.3 MW of neutral beam heating and a plasma current of 900 kA [14]. The inter-ELM pedestal electron density and temperature measured using Thomson laser scattering was approximately $5 \times 10^{19}$m$^{-3}$ and 200 eV respectively. The RFEA was reciprocated into the plasma and ion current measurements were made as a function of radial position as grid 1 of the RFEA was swept up to 600 V at a frequency of 2 kHz. The results from four repeat shots were combined to obtain a more comprehensive data set. Due to damage sustained to the RFEA graphite shell on the side facing the lower divertor (facing into the parallel plasma flow), only data from the side facing toward the upper divertor could be analysed. Fig. 1 shows the ion saturation current ($J_{sat}$) measured using the RFEA slit plate as a function of distance from the LCFS. Also shown is a exponential fit to the data with a decay length of 1.2 cm. Moving outwards from the LCFS, the results show a large drop in $J_{sat}$ within the first 2 cm followed by a more gradual fall off over the next 8 cm. As $J_{sat}$ is dependent on both density and temperature, its fall off with increasing distance from the plasma also shows the fall off in these quantities.

An example of inter-ELM data obtained 1.8 cm from the LCFS is shown in fig. 2. The collector measurements from two grid 1 sweeps were combined and binned and fitted using eq. (1) giving a $T_i$ of 68 eV. Ion temperature measured during the inter-ELM period is shown in fig. 3 as a function of distance from the LCFS. Also shown are $T_e$ measurements made using Thomson scattering (TS) [15] and $T_i$ measurements in the plasma core obtained using charge exchange recombination spectroscopy (CXRS) [16]. Linear (dotted line) and exponential (dashed line with a fall off length of 5 cm) fits to the CXRS measurements in the pedestal are also shown and act as a guide to what may be expected in the SOL. The RFEA data show that in the SOL, $T_i/T_e$ is approximately 2 to 6 with a mean value of 4. While the $T_e$ profile is relatively flat in the SOL, the $T_i$ profile is consistent with a fall off over the 2 to 10 cm region from the LCFS. The fall off of $T_i$ with distance from the LCFS lies between the linear and exponential extrapolations of the CXRS data. Previous measurements using CXRS [17] in a high collisionality plasma ($\upsilon^* > 1$) have shown $T_i \approx T_e$ in the pedestal region. Measurements in a plasma of similar collisionality to the one studied in this work ($\upsilon^* \approx 0.5$) have measured a flat $T_i$ profile in the pedestal region and extending out beyond the LCFS with $T_i \approx T_{e(pedestal)} = 300$ eV. Comparison of RFEA measurements in this work with those made by CXRS in reference [17] show a lower value of $T_i$ in the region beyond the LCFS where the two measurements overlap. . Investigation of the reason for this difference will form the basis for future work.

**3.2 ELM Measurements**

ELM measurements were made in both a type I and type III ELMing discharge. The type I ELMing discharge was a lower single null plasma with 1.9 MW of neutral beam heating and a plasma current of 650 kA. The pedestal electron density and temperature measured using Thomson scattering during the H-mode periods was approximately $3\times10^{19} m^{-3}$ and 200 eV respectively. The type III ELMing discharge was a connected double null plasma with 1.9 MW of beam heating and a plasma current of 600 kA. The pedestal electron density and temperature measured using Thomson scattering were approximately $2\times10^{19} m^{-3}$ and 100 eV respectively.

The filaments ejected into the SOL during ELMs which are detected by the RP, occur on a timescale of the order of tens of microseconds [18] which is less than the fastest sweep time of grid 1 (approximately 100 µs) and hence the ion $T_i$ of a single ELM cannot be measured. Instead, to determine the average $T_i$ in an ELM, slow sweeps of the grid 1 voltage were made (at a frequency of 20 Hz) in a method similar that used for ELM $T_i$ measurements on ASDEX Upgrade [10]. Due to the filamentry nature of ELMs, not all ELMs generated by the plasma will strike the RFEA. The plasma Dα emission trace was used to detect when an ELM occurred and the ELM was determined to have struck the RFEA if a spike was seen on the RFEA slit plate current trace during the ELM. When an ELM was detected by the RFEA, the collector current and grid 1 voltage at the time of the ELM were recorded. The accumulation of these measurements from a large number of ELMs allows graphs of ion current versus grid one voltage to be made at different distances from the LCFS. From these graphs, the ion temperature can be determined using equation (1). For both the type I and type III ELMing discharges only data from the side of the RFEA pointing away from the plasma flow is presented.

The results of ion energy measurements made during ELMs in a type III ELMing discharge at a distance of 3 to 17 cm from the LCFS are shown in fig. 4. Data from two repeat shots were combined to provide a larger data set. Ion saturation current measurements in fig. 4a show a fall off length of approximately 5 cm. Fig. 4b shows the collector current measured during the ELMs as a function of grid 1 voltage. The closed circles correspond to measurements made 3 to 6 cm from the LCFS and are shown along with a fit using equation (1) with a $T_i$=50 eV. The open circles correspond to measurements made 11 to 16 cm from the LCFS and are shown along with a fit using equation (1) with a $T_i$=30 eV. The error associated with fitting the data using equation (1) was approximately ± 10 eV.

Collector current measurements made during type I ELMs at a distance of 10 to 15 cm from the LCFS are shown in fig. 5 as a function of grid 1 voltage. For the type I ELMs, data from five repeat shots was accumulated to provide a larger data set. A fit to the measured data using equation (1) is also shown and gives a $T_i$ of 40 eV which is greater than measured in the type III ELMing discharge at a similar distance from the LCFS. The error associated with fitting the data using equation (1) was approximately ± 10 eV. The temperature of an ELM leaving the plasma is determined by the pedestal temperature and the radial velocity of the ELM. The radial velocity determines the extent of cooling the ELM will undergo before reaching a specific point. In MAST, Thomson laser scattering measurements of the pedestal $T_e$ show temperatures greater than 200 eV in type I ELMing discharges and $T_e$ of approximately 140 eV in type III ELMing discharges. Assuming $T_i$=$T_e$ at the pedestal, the lower pedestal temperature in type III ELMing discharges compared to type I ELMing discharges may account for the lower $T_i$ measured by the RFEA in type III ELMs compared to type I ELMs. The $T_i$ value measured during type I ELMs is approximately 20 % of the pedestal temperature which is lower than measurements made during type I ELMs in JET [3] and ASDEX-Upgrade [10]. However, these other measurements were made closer to the LCFS (approximately 3.5 to 8 cm from the LCFS) where higher ion temperatures would be expected.

## 4. Conclusion

Ion temperature measurements have been made in MAST using a midplane RFEA in H-mode plasmas during the inter-ELM period and during ELMs as a function of distance from the LCFS. During the inter-ELM period, ion temperatures of 10 to 70 eV were measured at distances of 3 to 8 cm from the LCFS. In the inter-ELM period in the SOL, $T_i/T_e$ was found to be 2 to 6 with a mean of 4. Measurements during type III ELMs gave $T_i \approx 50$ eV, 3 to 6 cm from the LCFS, decreasing to give $T_i \approx 30$ eV, 11 to 16 cm from the LCFS. Measurements during type I ELMs give $T_i \approx 40$ eV at a distance of 10 to 15 cm from the LCFS. The results obtained show that the ions in both type I and type III ELMs can carry significant energy into the far SOL and this qualitatively supports the results obtained in other machines during type I ELMs.

**Acknowledgements**

This work was part-funded by the RCUK Energy Programme under grant EP/I501045 and the European Communities under the contract of Association between EURATOM and CCFE. The views and opinions expressed herein do not necessarily reflect those of the European Commission.

**Figure Captions**

Fig. 1. Ion saturation current ($J_{sat}$) as a function of distance from the LCFS during the inter-ELM period of a 900 kA H-mode plasma. The different coloured/shaped points represent data from different repeat shots. Also shown is an exponential fit (fall off length of 1.2 cm) to the data.

Fig. 2. A typical collector current versus grid 1 voltage graph for the RFEA obtained during inter-ELM measurements. The results of two grid 1 voltage sweeps have been combined and binned (square points) and fitted using eq. (1).

Fig. 3. $T_i$ measurements made using the RFEA compared with $T_e$ measurements made using Thomson scattering (TS) and $T_i$ measurements made using charge exchange recombination spectroscopy (CXRS) as a function of radius during the inter-ELM period. The dotted and dashed lines are linear and exponential (fall off length 5 cm) fits to the CXRS $T_i$ data.

Fig. 4. Type III ELMing discharge results obtained using the RFEA. (a) Ion saturation current measurements made using the RFEA slit plate during ELMs as a function of distance from the LCFS. (b) Collector current measured during ELMs as a function of grid 1 voltage. The closed circles represent points measured 3 to 6 cm from the LCFS and are shown fitted using equation (1) with a $T_i$=50 eV. The open circles represent points measured 11 to 16 cm from the LCFS and are shown fitted using equation (1) with a $T_i$=30 eV.

Fig. 5. Collector current measured during type I ELMs as a function of grid 1 voltage. A fit using equation (1) with a $T_i$=40 eV is also shown.

**Figures**

Fig. 1

*Figure: $J_{SAT}$ (Am$^{-2}$) vs $\Delta R_{LCFS}$ (m) for shots 27504, 27505, 27601, 27602 with exponential fit.*

Fig. 2

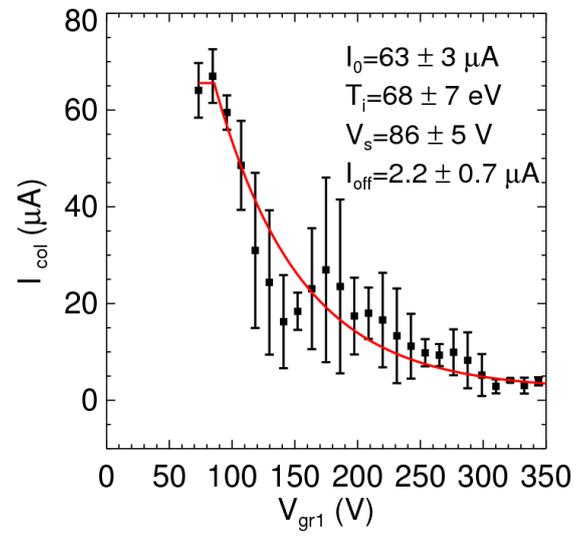

Fig. 3

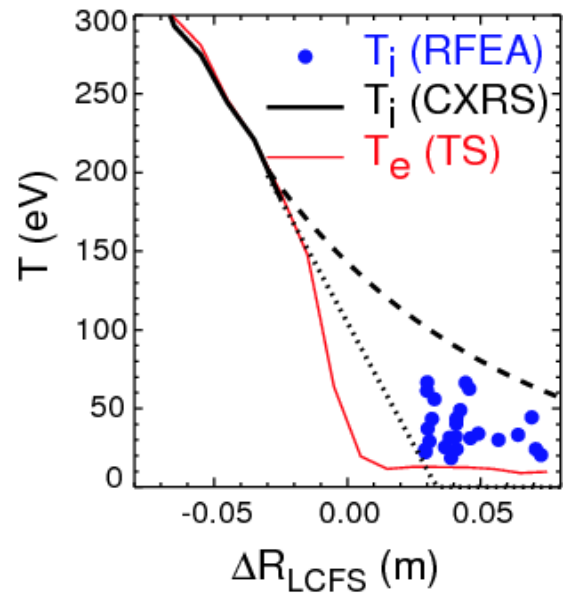

Fig. 4

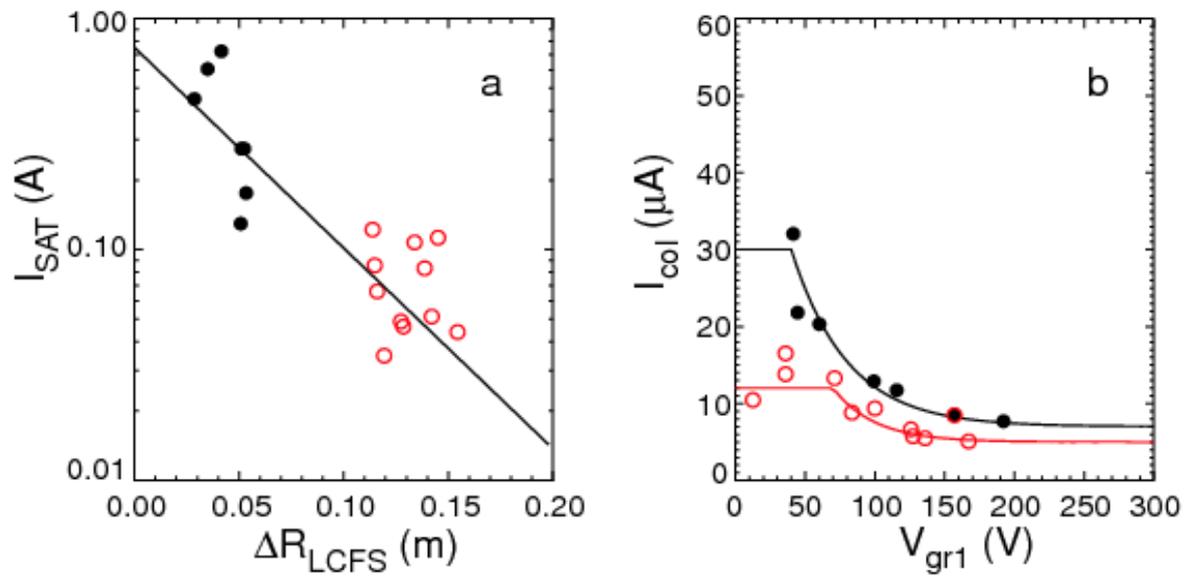

Fig. 5

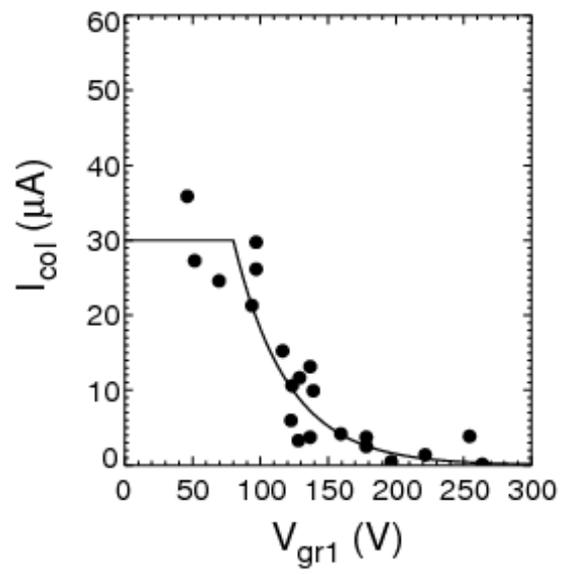